\documentclass[12pt]{iopart}
\usepackage[utf8]{inputenc}
\usepackage{physics}

\usepackage{color}
\usepackage{subcaption}
\usepackage{amsmath}
\usepackage{amsfonts} 
\usepackage{hyperref}
\usepackage{mathrsfs}
\usepackage{tikz}   
\usetikzlibrary{shapes,arrows,positioning,automata,backgrounds,calc,er,patterns}
\usepackage{tikz-feynman}
\tikzfeynmanset{compat=1.0.0}

\begin{document}
\title[Fluctuating hydrodynamics of aligning active particles: the dilute limit]{Fluctuating kinetic theory and fluctuating hydrodynamics of aligning active particles: the dilute limit}
\author{Ouassim Feliachi$^1$, Marc Besse$^2$,
Cesare Nardini$^{2,3}$\ and Julien Barré$^1$}
\address{$^1$ Institut Denis Poisson, Université d’Orléans, Université de Tours and CNRS, 45067 Orléans, France}
\address{$^2$ Sorbonne Universit\'e, CNRS, Laboratoire de Physique Th\'eorique de la Mati\`ere Condens\'ee, 75005 Paris, France}
\address{$^3$ Service de Physique de l'Etat Condens\'e, CEA, CNRS Universit\'e Paris-Saclay, CEA-Saclay, 91191 Gif-sur-Yvette, France}
\ead{ouassim.feliachi@univ-orleans.fr}

\begin{abstract}
Kinetic and hydrodynamic theories are widely employed for describing the collective behaviour of active matter systems. At the fluctuating level, these have been obtained from explicit coarse-graining procedures in the limit where each particle interacts weakly with many others, so that the total forces and torques exerted on each of them is of order unity at all times. Such limit is however not relevant for dilute systems that mostly interact via alignment; there, collisions are rare and make the self-propulsion direction to change abruptly. We derive a fluctuating kinetic theory, and the corresponding fluctuating hydrodynamics, for aligning self-propelled particles in the limit of dilute systems. We discover that fluctuations at kinetic level are not Gaussian and depend on the interactions among particles, but that only their Gaussian part survives in the hydrodynamic limit. At variance with fluctuating hydrodynamics for weakly interacting particles, we find that the noise variance at hydrodynamic level depends on the interaction rules among particles and is proportional to the square of the density, reflecting the binary nature of the aligning process. The results of this paper, which are derived for polar self-propelled particles with polar alignment, could be straightforwardly extended to polar particles with nematic alignment or to fully nematic systems.
\end{abstract}

\submitto{\JSTAT}
\maketitle

\section{Introduction}
Active systems are composed of units able to extract non-thermal energy from the environment and dissipate it to self-propel~\cite{marchetti2013hydrodynamics}. Examples span a broad range of scales, from bacteria to animals in the biological world, and significant effort has been devoted recently to build synthetic active systems in the laboratory using self-propelled granular particles~\cite{deseigne2010collective}, Janus particles~\cite{palacci2013living} or Quincke rollers~\cite{bricard2013emergence} to name just a few examples. Active systems break detailed balance microscopically, as opposed to more classical non-equilibrium systems where detailed balance is broken by boundary driving. As such, they are capable of novel collective behaviours that are impossible in equilibrium systems, whose characterization and control attracted a significant attention recently~\cite{marchetti2013hydrodynamics}. When the dominant interaction among particles is to align their direction of motion, which can be caused by collision when particles have anisotropic shape, or by reaction to sensing, a well-known collective behavior of active systems emerges: flocking. This is a ferromagnetic-like state where all particles move in average along a given direction; broken detailed balance allows for long range order even in two-dimensions and a scale-free structure giving rise to long-range correlations without the need of fine-tuning to criticality~\cite{Vicsek1995,toner1995long}. 


One of the main tools used to investigate the collective behavior of active systems are fluctuating hydrodynamic theories. These are field theories that retain only the slow fields to describe the large-scale, long-time evolution of the system and can be derived via two complementary paths. On one hand, they can be written on the basis of symmetry arguments~\cite{marchetti2013hydrodynamics,toner2005hydrodynamics,cates2018theories}; this approach is particularly useful for studying active systems, given that their complexity often does not allow to build first-principle models even at the microscopic level; it had great success to unveil new generic and universal (i.e. qualitative and quantitative properties independent of system details) physics induced by activity. It has two shortcomings though: first, symmetries do not allow to relate microscopic parameters to those entering in the fluctuating hydrodynamics. Second, in active systems the noise term is not constrained by the fluctuation-dissipation theorem, and it is unclear how to specify it a-priori, except when dealing with critical systems (cases in which Renormalization Group arguments allow to discard irrelevant nonlinearities). It should be noted that this feature is at variance not only with equilibrium systems, but also with non-equilibrium ones weakly driven by the boundaries; in these, at least for weak coupling, the noise term is constrained by linear response theory~\cite{bertini2015macroscopic}. Hence there is something to learn from linking the microscopic and macroscopic descriptions of active systems even if the starting point are phenomenological particle models often chosen only on the basis of simplicity, and several works in the literature have been indeed focused on this program~\cite{Bertin2006,Bertin2009,degond2008continuum,chate2019dry}. 


Kinetic theories and hydrodynamic limits form a backbone of classical statistical mechanics, allowing to connect microscopic models to their long-time, large-scale description~\cite{balescu1997statistical}. Broadly speaking, controlled kinetic theory description can be derived in two opposite limits: when the system is very dilute, the classical Botzmann-Grad limit where the Boltzmann equation is derived~\cite{cercignani1988boltzmann,balescu1997statistical}, or when each particle interacts weakly with many others so that the typical force exerted on each of them is of order unity~\cite{villani2002review,balescu1997statistical}. This latter class comprises systems with long-range interactions (plasmas or self-gravitating systems)~\cite{balescu1997statistical,bouchet2010thermodynamics} and polymers~\cite{doi1988theory}. Kinetic descriptions, describing the evolution of the one-particle distribution function in phase space, is often too complex to be studied either analytically or numerically. The hydrodynamic limit is then often employed, in which only the slow macroscopic fields, such as the density and momenta of particles are retained in the description. A classical example is the derivation of the Navier-Stokes equation from the Boltzmann equation via the Chapman--Enskog expansion in the small Knudsen number limit $\alpha = \ell/L\ll 1$, where $\ell$ is the mean free path of particles and $L$ a macroscopic length scale~\cite{bardos1991fluid}. 

Classically, kinetic and hydrodynamic theories have been developed at the `mean-field' level, i.e. discarding fluctuations at the large-scales. The derivation of kinetic and hydrodynamic theories at the fluctuating level, properly deriving the noise term that induces the fluctuations of the relevant mesoscopic fields, has seen significant developments in the last 30 years. One of the most widely employed methods was initially developed by Dean~\cite{Dean_1996} and Kawasaki~\cite{kawasaki1994stochastic} to describe overdamped diffusing particles, whose formal derivation can be precisely justified in the limit of weak interactions among particles~\cite{bouchet2016perturbative,barre2015motility}. In this approach, the noise at hydrodynamic level is independent of the particles interactions, and equal to the one of freely diffusing particles.  Recently, a technique to derive the fluctuating kinetic theory of perfect gases -- in the Boltzmann-Grad limit, was introduced in the mathematical~\cite{rezakhanlou1998large} and physics~\cite{Bouchet2020} literature. So far, however, no derivation of the ensuing fluctuating hydrodynamics has been proposed\footnote{Work in progress by J. Barr\'e, F. Bouchet and O. Feliachi.}.

Kinetic and hydrodynamic theories have been widely employed for describing systems of self-propelled particles interacting via alignment. This route has indeed been followed both within the weak-interactions limit~\cite{degond2008continuum} and within the Boltzmann-like framework of dilute systems~\cite{Bertin2006,Bertin2009,chate2019dry}. The Dean-Kawasaki approach has been  widely employed to derive the fluctuating kinetic theory and fluctuating hydrodynamics of microscopic active matter models. This is justified when interactions are long-ranged, as it happens for dilute microswimmer suspensions in which the primary source of interactions are low-Reynolds fluid flows created by the motion of the swimmers~\cite{qian2017stochastic,stenhammar2017role,vskultety2020swimming}. Yet, the fact that hydrodynamics noise is independent of interactions within the Dean approach motivated some authors to use it even for short-ranged aligning particles~\cite{bertin2013mesoscopic}, even if these systems are clearly out of the regime of applicability of the method. For dilute systems, indeed, although particle diffusion will give rise to a Dean-like noise, one can expect another contribution from particle-particle collisions.


In this paper, we describe how to derive the fluctuating kinetic theory and the corresponding fluctuating hydrodynamics of active particles that interact by aligning. The fluctuating kinetic theory is obtained in the dilute limit, analogous to the Boltzmann-Grad limit of perfect gases. This leads to a noise term at kinetic level that is not Gaussian. We then derive the corresponding fluctuating hydrodynamics in two limits. First, deeply in the ordered phase so that the slow fields are the density and the local orientation of the polar order. This result extends to the dilute limit, both at the deterministic and fluctuating levels, previous results that have been obtained in the weak-interactions limit~\cite{degond2008continuum}, and it is valid when the Knudsen number is small. Second, we derive the fluctuating hydrodynamics close to the order-disorder transition, extending at the fluctuating level the deterministic hydrodynamic theory developed in~\cite{Bertin2006,Bertin2009}. Interestingly, the noise entering at the fluctuating level is Gaussian, and we explicitly compute its variance. The latter turns out to be proportional to the square of the density field and to depend explicitly on the interactions among particles; both these facts differentiate our conclusion from the results obtained in the Dean-Kawasaki approach, where the noise variance is linear in the density and independent from particle-particle interactions~\cite{bertin2013mesoscopic}. 

The paper is organized as follows. In section \ref{sec:KT} we specify the particle-based model we consider and, under an extended molecular chaos type hypothesis, derive its kinetic theory and the associated fluctuating kinetic theory, described as a dynamical Large Deviation Principle (LDP). As a warm-up problem we first derive it for independent Run-and-Tumble particles in section \ref{subsec:RT}, and then for the interacting case in section \ref{subsec:LDP-Boltzmann}. The large deviation rate function we obtain is not quadratic, which corresponds to a fluctuating kinetic theory with a non Gaussian noise. In section \ref{sec:degondderiv}, we start from the fluctuating kinetic theory to derive fluctuating hydrodynamic equations at leading order in the Knudsen number $\alpha$, deeply in the ordered phase. In particular, we show that in this limit $\alpha\to 0$, the noise becomes Gaussian. Finally, in section \ref{sec:hydrophasetransition}, we start again from the fluctuating kinetic theory and present the derivation of the fluctuating hydrodynamics close to the order-disorder transition. We show that also in this case the noise becomes Gaussian, and discuss the links with section \ref{sec:degondderiv}.


\section{Definition of the particle-based model, kinetic theory and dynamical large deviations}
\label{sec:KT}

We start by introducing the particle-based model we consider, that we term the Boltzmann--Vicsek particle model, in section \ref{subsec:Pmodel}; in section \ref{subsec:Kineticeq} we describe its well-known kinetic description at the deterministic level (known as Boltzmann--Vicsek equation). We then introduce a suited non-dimensional system of units that allows to investigate fluctuations at the kinetic level in section \ref{subsec:rescaled-boltzmann}. We discuss the LDP for run and tumble particles in section \ref{subsec:RT} and then, adapting the arguments of \cite{Bouchet2020}, we derive in section \ref{subsec:LDP-Boltzmann} the fluctuating kinetic theory associated with the particle-based model as a large deviation principle. Some of the properties of the fluctuating Boltzmann-Vicsek equation are discussed in section \ref{subsec:Some-properties-of}.

\subsection{Boltzmann--Vicsek particle model}

\label{subsec:Pmodel}

We consider $N$ particles evolving in a periodic
two-dimensional box of size $L\times L$. We denote $\left(\mathbf{r}_{n},\theta_{n}\right)_{1\leq n\leq N}$
their positions and orientations according to some arbitrary axis. The dynamics is the one first introduced in \cite{Bertin2006}. Particles
move ballistically with constant speed $v_{0}$: $d\textbf{r}_{i}/dt=v_{0}(\cos{\theta_{i}},\sin{\theta_{i}})$,
until they collide. When two particles $i$ and $j$ are close enough (i.e. $\left|\mathbf{r}_{i}-\mathbf{r}_{j}\right|\leq2R$, $R$ being the interaction radius) a collision occurs with a rate $(v_{0}/R)K(\theta_{i}-\theta_{j})$ where $K$ is a cross-section chosen to mimic hard-sphere collisions. This rate is furthermore chosen so that when two particles meet, they have a probability to interact of order $1$. When a collision occurs, particles update their orientation according to the following rule 
\[
\theta_{i}^{\text{out}}=\bar{\theta}+\zeta_{i},\,\,\,\theta_{j}^{\text{out}}=\bar{\theta}+\zeta_{j},
\]
where $\bar{\theta}=\arg\left(\text{e}^{i\theta_{i}^{\text{in}}}+\text{e}^{i\theta_{j}^{\text{in}}}\right)$ and
the superscript ``in'' (resp. ``out'') denotes incoming (resp. outcoming) orientations. $\zeta_{i}$ and $\zeta_{j}$ are independent random
variables distributed according to $P_{\sigma}(\theta)$
over $[-\pi,\pi)$ with variance $\sigma^{2}$. At low variance of the noise, this interaction favors the polar alignment of particles. 

\begin{figure}[!t]
\centering
\begin{tikzpicture}[line cap=round,line join=round,>=triangle 45,x=1.0cm,y=1.0cm]
\clip(-5.3,-3.3) rectangle (6.8,3.3);
\draw [fill=black,fill opacity=1.0] (-4,2) circle (0.25cm);
\draw [fill=black,fill opacity=1.0] (-4,-2) circle (0.25cm);
\draw [fill=black,fill opacity=1.0] (4,1) circle (0.25cm);
\draw [fill=black,fill opacity=1.0] (4,-1) circle (0.25cm);
\draw [dash pattern=on 6pt off 6pt] (-4,2) circle (1cm);
\draw [dash pattern=on 6pt off 6pt] (-4,-2) circle (1cm);
\draw [dash pattern=on 6pt off 6pt] (4,1) circle (1cm);
\draw [dash pattern=on 6pt off 6pt] (4,-1) circle (1cm);
\draw [dotted] (-10,0)-- (7,0) ;
\draw [dotted] (-4,2)-- (-2.04,1.51);
\draw [shift={(-4,2)},dotted]  plot[domain=-0.24:0,variable=\t]({1*2.02*cos(\t r)+0*2.02*sin(\t r)},{0*2.02*cos(\t r)+1*2.02*sin(\t r)});
\draw [shift={(-4,-2)},dotted]  plot[domain=0:0.24,variable=\t]({1*2.04*cos(\t r)+0*2.04*sin(\t r)},{0*2.04*cos(\t r)+1*2.04*sin(\t r)});
\draw [shift={(4,-1)},dotted]  plot[domain=-0.8:0,variable=\t]({1*1.88*cos(\t r)+0*1.88*sin(\t r)},{0*1.88*cos(\t r)+1*1.88*sin(\t r)});
\draw [shift={(4,1)},dotted]  plot[domain=0:0.14,variable=\t]({1*1.88*cos(\t r)+0*1.88*sin(\t r)},{0*1.88*cos(\t r)+1*1.88*sin(\t r)});
\draw [dotted] (-4,2)-- (-2,2);
\draw [dotted] (-4,-2)-- (-1.96,-2);
\draw [dotted] (-4,-2)-- (-2.05,-1.51);
\draw [dotted] (4,1)-- (5.88,1);
\draw [dotted] (4,1)-- (5.89,1.27);
\draw [dotted] (4,-1)-- (5.31,-2.35);
\draw [dotted] (4,-1)-- (5.88,-1);
\draw [->] (-4,-2) -- (-2.05,-1.51);
\draw [->] (-4,2) -- (-2.04,1.51);
\draw [->] (4,1) -- (5.89,1.27);
\draw [->] (4,-1) -- (5.31,-2.35);
\path (-1.9,1.78) node [right] {$\theta_1^{\rm in}$};
\path (-1.88,-1.76) node [right] {$\theta_2^{\rm in}$};
\path (5.8,-1.7) node [right] {$\theta_2^{\rm out}$};
\path (5.9,1.2) node [right] {$\theta_1^{\rm out}$};
\end{tikzpicture}
\caption{Schematic representation of a collision event. In this specific case, $\bar{\theta}=\arg\left(\text{e}^{i\theta_{1}^{\text{in}}}+\text{e}^{i\theta_{2}^{\text{in}}}\right)=0$ with respect to the dotted axis, $\zeta_1 = \theta_{1}^{\text{out}}$, and $\zeta_2 = \theta_{2}^{\text{out}}$.}
\label{fig:collision_inelastique}
\end{figure}
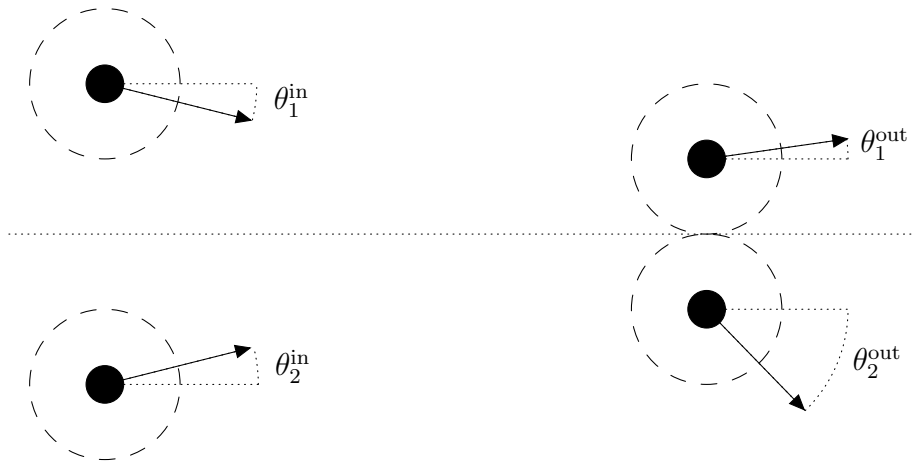

It should be observed that in the model, at variance with the standard Vicsek model that is often considered in computational works~\cite{Vicsek1995,chate2020dry}, only binary collisions are considered. The collision process is schematically presented in figure \ref{fig:collision_inelastique}. In the following, this model is called the Boltzmann--Vicsek particle model.

\subsection{Boltzmann-Vicsek equation}

\label{subsec:Kineticeq}

The deterministic kinetic description associated with the Boltzmann--Vicsek particle model was derived in \cite{Bertin2006} and reads  
\begin{equation}
\partial_{t}f_e(\mathbf{r},\theta,t)+v_{0}\mathbf{e}_{\theta}\cdot\nabla f_e(\mathbf{r},\theta,t)=v_{0}R\mathcal{I}_{\rm col}[f_e](\mathbf{r},\theta,t)\label{eq:BoltzmannVicsek_dim}
\end{equation}
where $f_e( \mathbf{r},\theta ,t)$ is the one-particle distribution function in the phase-space (representing the number of particles at a position $\mathbf{r}$, with orientation $\theta$ at a certain time $t$) normalised such that $\int d \mathbf{r}d\theta \, f_e = N$. In (\ref{eq:BoltzmannVicsek_dim}) the collision term is given by
\begin{align}
\label{eq:I-coll}
& \mathcal{I}_{\rm col}[f_e](\mathbf{r},\theta,t)=\iint d\theta_{1}d\theta_{2}\,f_e(\mathbf{r},\theta_{1},t)f_e(\mathbf{r},\theta_{2},t)K(\theta_{2}-\theta_{1}) \\
& \times \left\{ P_{\sigma}(\theta-\Psi(\theta_{1},\theta_{2}))-\delta(\theta-\theta_{1})\right\}, \nonumber \\
& K(\theta_{2}-\theta_{1})=2\abs{\sin{\bigg(\frac{\theta_{2}-\theta_{1}}{2}}\bigg)} \ \text{is the scattering cross-section,} \\
& \Psi(\theta_{1},\theta_{2})=\arg{\big(e^{i\theta_{1}}+e^{i\theta_{2}}\big)}=\theta_{1}+H(\theta_{2}-\theta_{1}) \label{eq:psi} \\
& H(\Delta)=\frac{\Delta}{2}, \forall\Delta\in\ [-\pi,\pi),~H \text{ is } 2\pi{\rm -periodic}.
\end{align}
The Boltzmann--Vicsek equation \eqref{eq:BoltzmannVicsek_dim} relies on the molecular chaos hypothesis and it is expected to be a valid description of the particle system in the limit of a large number of particles in the Boltzmann-Grad limit, as is made
explicit in the next section. 

\subsection{The rescaled Boltzmann--Vicsek equation}

\label{subsec:rescaled-boltzmann}

We introduce a set of units that are suited to investigate the kinetic limit: space is measured in units of the mean free path $\ell  = 1/(R\rho_0)$, where $\rho_{0}=N/L{^{2}}$ is the
mean density, and time in units of $\ell/v_{0}$, which is the average time between two collisions. We also define $\epsilon=\left(\rho_{0}\ell^{2} \right)^{-1}$, the inverse of the number of particles
in a region of surface $\ell^2$. By performing a space-time rescaling $\mathbf{r}'=\mathbf{r}/\ell,$ $t'=tv_{0}/\ell$ and by rescaling the distribution function $f(\mathbf{r}',\theta,t')=\epsilon f_e(\mathbf{r}',\theta,t')$ (primes are dropped afterwards), the Boltzmann--Vicsek equation reads 
\begin{equation}
\partial_{t}f+\mathbf{e}_{\theta}\cdot\nabla f=\mathcal{I}_{\rm col}[f].\label{eq:rescaledboltzmannvicsek}
\end{equation}
As we shall see below, the Boltzmann--Vicsek equation is a valid description of the
microscopic model in the limit $N\to +\infty$, $\epsilon\to0$ (and under the molecular chaos hypothesis). It should be noticed that $\epsilon=R/\ell=NR^2/L^2$, meaning that the limit yielding a Boltzmann-type kinetic description is opposite to a weak-interaction limit for which the number of particles in an interaction radius goes to infinity. 

In the next section, we go beyond this law of large
numbers, taking into account fluctuations by determining the LDP for the empirical distribution. Before addressing the derivation of a LDP for the Boltzmann--Vicsek equation (i.e. the fluctuating kinetic theory), we show in the next section how the LDP can be obtained for non-interacting Run-and-Tumble particles; this minimal problem bears technical similarities with the problem we address and it is thus instructive to consider it first.

\subsection{Large deviations for the empirical distribution of $N$ Run-and-Tumbling particles}
\label{subsec:RT}

In this section, we derive a LDP that describes
the probability for an evolution path of the empirical distribution
$f_{N}(\mathbf{r},\theta,t)=N^{-1}\sum_{n}\delta\left(\mathbf{r}{}_{n}-\mathbf{r}\right)\delta\left(\theta_{n}-\theta\right)$
of $N$ non-interacting particles undergoing a Run-and-Tumble dynamics to be close to the evolution path of a prescribed smooth distribution $f$.
The LDP reads
\begin{equation}
\mathbb{P}\left[\left\{ f_{N}(t)\right\} _{0\leq t<T}=\left\{ f(t)\right\} _{0\leq t<T}\right]\underset{N\uparrow\infty}{\asymp}\exp\left(-N\int_{0}^{T}\mbox{d}t\,\sup_{p}\left(\int d\mathbf{r}d\theta\,p\partial_{t}f-H_{RT}\left[f,p\right]\right)\right),\label{eq:LDP-RT}
\end{equation}
where $p(\mathbf{r},\theta)$ is the ``momentum'' conjugated to $\partial_t f$, the symbol $\underset{N\uparrow \infty}{\asymp}$ is the logarithm equivalence
\begin{equation}
\varphi_{N}\underset{N\uparrow \infty}{\asymp}\exp(N \psi )\iff\lim_{N\uparrow \infty}N^{-1} \log \varphi_{N}=\psi
\end{equation}
and $H_{RT}$ is called the large deviation Hamiltonian, a
functional of both $f$ and $p$. $H_{RT}$ encodes
all the dynamical statistical properties of the empirical distribution
$f_{N}$. In section \ref{subsec:Runandtumble-dynamics}, we introduce the Run-and-Tumble particle dynamics. In section \ref{subsec:LDP-R-T},
we explain how to compute $H_{RT}$ in the case of Run-and-Tumble
particles. 

\subsubsection{Particle dynamics and kinetic description.\label{subsec:Runandtumble-dynamics}}

We consider $N$ particles in a two-dimensional periodic domain traveling at a constant speed $v_{0}$. A particle changes its orientation from $\theta$ to $\theta'$ with a rate $\lambda$ following a  distribution $P_{t}$
on $[0,2\pi)$ which is symmetric with respect to $\theta$. At the kinetic level the distribution function $f(\mathbf{r},\theta,t)$
of the position and orientation of the $N$ particles satisfies
\[
\partial_{t}f+v_{0}\mathbf{e}_{\theta}\cdot\nabla f=-\lambda f+\lambda\int\text{d}\theta'\,P_{t}\left(\theta'-\theta\right)f\left(\theta'\right),
\]
where $f$ is normalized to 1.
We define the mean free path for the Run-and-Tumble particles $\ell=v_{0}/\lambda$ and we rescale space and time $\mathbf{r}'=\mathbf{r}/\ell, t'=tv_{0}/\ell$. Dropping the primes, it yields
\begin{equation}
\partial_{t}f+\mathbf{e}_{\theta}\cdot\nabla f=-f+\int\text{d}\theta'\,P_{t}\left(\theta'-\theta\right)f\left(\theta'\right).\label{eq:RT-kinetics}
\end{equation}
Equation 
\eqref{eq:RT-kinetics} can be seen as a law of large numbers
for the empirical distribution $f_{N}$: in the limit of a large number of particles, the random object $f_N$ concentrates
on the distribution function $f$ which is a solution of \eqref{eq:RT-kinetics}. 

\subsubsection{Large deviations for the empirical distribution\label{subsec:LDP-R-T}.}
We now assess the probability
of any evolution path for the empirical distribution. As shown
in \cite{freidlin1998random}, a way to compute the large
deviation Hamiltonian $H$ associated with \eqref{eq:RT-kinetics} is to compute
the infinitesimal generator of the Markov process describing the evolution
of the empirical distribution $f_{N}$. Then, from the infinitesimal
generator $G_{f}$, the large deviation Hamiltonian is deduced through
the following formula
\begin{equation}
H_{RT}[f,p]=\lim_{N\uparrow\infty}\frac{1}{N}G_{f}\left[\mbox{e}^{N\int d\mathbf{r}d\theta\,pf_{N}}\right]\mbox{e}^{-N\int d\mathbf{r}d\theta\,pf},\label{eq:HamiltoGen}
\end{equation}
where the definition of the infinitesimal generator is
\begin{equation}
G_{f}\left[\phi\right]=\lim_{t\rightarrow0}\frac{\mathbb{E}_{f}\left[\phi\left[f_{N}\left(t\right)\right]\right]-\phi\left[f\right]}{t},\label{eq:generator}
\end{equation}
where $\phi$ is a test functional of the empirical distribution.
In \eqref{eq:generator}, $\mathbb{E}_{f}$ denotes an expectation
over the stochastic process $f_{N}$ conditioned by $f_{N}(t=0)=f$.
The generator can be split into two terms
\[
G_{f}=G_{f,\mathcal{T}}+G_{f,{\rm tumb}},
\]
where $G_{f,\mathcal{T}}$ is due to free transport, and $G_{f,{\rm tumb}}$ to tumbling
events. A Taylor expansion of $\phi\left[f_{N}\left(t\right)\right]$ at small times
allows to compute the transport part of the generator
\begin{equation}
G_{f,\mathcal{T}}[\phi]=-\int d\mathbf{r}d\theta\,\mathbf{e}_{\theta}\cdot\nabla f\frac{\delta\phi}{\delta f\left(\mathbf{r},\theta\right)}.\label{eq:G_T}
\end{equation}
To compute $G_{f,{\rm tumb}}$, we need to evaluate the effect of tumbling
events on the empirical distribution. If $f$ is the empirical
distribution, the rate of tumbling events that change the orientation
of a particle from $\theta_{1}$ to $\theta_{1}^{'}$ in
the volume element $d\mathbf{r}_{1}$ centered at point $\mathbf{r}_{1}$
is: 
\begin{equation}
Nf(\mathbf{r}_{1},\theta_{1},t)P_{t}\left(\theta_{1}-\theta'_{1}\right)d\theta_{1}d\theta_{1}^{'}d\mathbf{r}_{1}.\label{eq:changeempRT}
\end{equation}
Each tumbling event of this type changes the empirical distribution
from $f\left(\mathbf{r},\theta\right)$ to $f\left(\mathbf{r},\theta\right)-N^{-1}\delta\left(\mathbf{r}-\mathbf{r}_{1}\right)\delta\left(\theta-\theta_{1}\right)+N^{-1}\delta\left(\mathbf{r}-\mathbf{r}_{1}\right)\delta\left(\theta-\theta'_{1}\right).$
Therefore, from \eqref{eq:generator} and \eqref{eq:changeempRT},
we deduce the part of the infinitesimal generator due to tumbling
events
\begin{equation}
\begin{split}
    & G_{f,{\rm tumb}}[\phi]=N\int d\theta_{1}d\theta'_{1}d\mathbf{r}\ f(\mathbf{r},\theta_{1},t)P_{t}\left(\theta_{1}-\theta'_{1}\right)  \left(\phi[\tilde{f}]-\phi[f]\right),
\end{split}
\end{equation}
where $\tilde{f}\left(\mathbf{r}_0,\theta,t\right)=f\left(\mathbf{r}_0,\theta,t\right)+N^{-1}\delta(\mathbf{r}_{0}-\mathbf{r})\left(-\delta(\theta-\theta_{1})+\delta(\theta-\theta'_{1})\right)$.
We can then apply \eqref{eq:HamiltoGen} to deduce the large deviation
Hamiltonian
\begin{equation}
H_{RT}\left[f,p\right]=H_{\mathcal{T}}\left[f,p\right]+H_{{\rm tumb}}\left[f,p\right],\label{eq:H_tumbling}
\end{equation}
where
\begin{align}
& H_{\mathcal{T}}[f,p] =-\int d\theta d\mathbf{r}\,p(\mathbf{r},\theta,t)\mathbf{e}_{\theta}\cdot\nabla f(\mathbf{r},\theta,t)
,\label{eq:H_T} \\
& H_{{\rm tumb}}\left[f,p\right]=\int d\theta_{1}d\theta'_{1}d\mathbf{r}\,f(\mathbf{r},\theta_{1},t)P_{t}\left(\theta_{1}-\theta'_{1}\right)\bigg\{ e^{-p(\mathbf{r},\theta_{1},t)+p(\mathbf{r},\theta'_{1},t)}-1\bigg\}.\label{eq:H_t}
\end{align}

The most probable evolution for the empirical distribution 
is the one that maximizes the right hand side of the LDP
\eqref{eq:LDP-RT}. This maximization condition is simply the Hamilton
equation associated with the large deviation Hamiltonian (\ref{eq:H_tumbling}), which gives:
$\partial_{t}f=\frac{\delta H_{RT}}{\delta p}\left[f,p=0\right]$ or, explicitely, equation (\ref{eq:RT-kinetics}).

Tumbling events conserve locally the number of particles. This  is shown by the fact that $\int d\mathbf{r}d\theta \partial_t f = \int d\mathbf{r}d\theta \,\delta H_{RT}/\delta p(\mathbf{r},\theta) =0$, where the first equality follows from the Hamilton's equations, and for the second we have used  (\ref{eq:H_t}).
The large deviation Hamiltonian $H_{RT}$ is non-quadratic in the
conjugated momentum $p$. This means that, if we wanted to write a stochastic partial differential equation for the empirical distribution, it would contain non-Gaussian noise.

\subsubsection{Time-reversibility.\label{subsec:Some-properties-of}}
In the absence of interactions, the quasipotential is given by Sanov's theorem~\cite{sanov1957probability}; the probability for the empirical distribution to be close to a certain distribution $f$ is given by the number of phase-space configurations that are compatible with this distribution $f$:
\begin{equation}
\mathbb{P}_{S}\left(f_{N}=f\right)\underset{N\uparrow\infty}{\asymp}\exp\left(NS\left[f\right]\right),\label{eq:Quasipot}
\end{equation}
where $S\left[f\right]=-\int d\theta d\mathbf{r}\,f\log f$ is the entropy. Indeed, a necessary condition for the compatibility of \eqref{eq:Quasipot} and the LDP \eqref{eq:LDP-RT}
is provided by the Hamilton--Jacobi equation: 
\begin{equation}
H_{RT}\left[f,-\frac{\delta S}{\delta f}\right]=0\label{eq:hamilton_jacobi}
\end{equation}
which can be explicitly checked to hold. This fact is related to the presence of the generalised time-reversal symmetry $\theta\to \theta+\pi,t\to -t$. Defining $\mathscr{S}[f](\mathbf{r},\theta,t)=f(\mathbf{r},\theta+\pi,-t)$, this symmetry translates into the following identity for the large deviation Hamiltonian \cite{Bouchet2020}:
\begin{equation}
H_{RT}\big[\mathscr{S}\left[f\right],-\mathscr{S}\left[p\right]\big]=H_{RT}\left[f,p-\frac{\delta S}{\delta f}\right] \label{eq:hamiltonian_detailed}.
\end{equation}
Time-reversal symmetry breaks down for the Boltzmann-Vicsek model and the solution of \eqref{eq:hamilton_jacobi}, which would play the role of the entropy, is unknown.

\subsection{Large deviations from the Boltzmann--Vicsek equation}

\label{subsec:LDP-Boltzmann}

We now aim at deriving the fluctuating kinetic theory associated with the microscopic model introduced in section \ref{subsec:Pmodel}, along the same lines as in section \ref{subsec:RT}. We expect a LDP for the rescaled empirical distribution 
\begin{align}
f_{\epsilon}\left(\mathbf{r},\theta,t\right)=\epsilon\sum_{n=1}^{N}\delta\left(\mathbf{r}_{n}\left(t\right)-\mathbf{r}\right)\delta\left(\theta_{n}\left(t\right)-\theta\right),
\end{align}
in the form
\begin{equation}
\mathbb{P}\left[\left\{ f_{\epsilon}(t)\right\} _{0\leq t<T}=\left\{ f(t)\right\} _{0\leq t<T}\right]\underset{\epsilon\downarrow0}{\asymp}\exp\left(-\frac{1}{\epsilon}J_T[f]\right),\label{eq:LDP_Boltzmann}
\end{equation}
where
\begin{align}
J_{T}[f]=\int_{0}^{T}dt\,\sup_{p}\left(\int d\mathbf{r}d\theta\,\partial_{t}fp-H_{BV}\left[f,p\right]\right), \\
H_{BV}\left[f,p\right]=\lim_{\epsilon\downarrow0}\epsilon G_{f}\left[\mbox{e}^{\frac{1}{\epsilon}\int d\mathbf{r}d\theta\,pf_{\epsilon}}\right]\mbox{e}^{-\frac{1}{\epsilon}\int d\mathbf{r}d\theta\,pf}.
\end{align}
In \eqref{eq:LDP_Boltzmann} and in every other equivalences that involve $\epsilon \to 0$, we also implicitly take the $N\to +\infty $ limit.

We start from the definition of the infinitesimal generator \eqref{eq:generator}.
This time, the expectation $\mathbb{E}_{f}$ denotes an expectation
over the stochastic process of the rescaled empirical distribution
$f_{\epsilon}$ of $N$ particles submitted to the Boltzmann--Vicsek
dynamics conditioned by $f_{\epsilon}(t=0)=f$. As previously, we
can decompose the infinitesimal generator in two terms
\[
G_{f}=G_{f,\mathcal{T}}+G_{f,{\rm col}},
\]
where $G_{f,\mathcal{T}}$ is the infinitesimal generator accounting for
free transport, already computed in \eqref{eq:G_T}, and $G_{f,{\rm col}}$ accounts for two-body
collisions. To evaluate $G_{f,{\rm col}}$, we need the rate of two-body collisions, which change the orientation
of two particles from $(\theta_{1},\theta_{2})$ to $(\theta'_{1},\theta'_{2})$
in the volume element $d\mathbf{r}$ centered at point $\mathbf{r}$. If $f$ is the rescaled empirical distribution, this rate reads
\begin{equation}
\frac{1}{2\epsilon}K(\theta_{2}-\theta_{1})f(\mathbf{r},\theta_{1},t)f(\mathbf{r},\theta_{2},t)P_{\sigma}\left(\theta'_{1}-\Psi(\theta_{1},\theta_{2})\right)P_{\sigma}\left(\theta'_{2}-\Psi(\theta_{1},\theta_{2})\right)d\theta_{1}d\theta_{2}d\theta'_{1}d\theta'_{2}d\mathbf{r}.\label{eq:changeempRT-1}
\end{equation}
As we did to justify the Boltzmann--Vicsek equation \eqref{eq:BoltzmannVicsek_dim}, we assumed the molecular chaos hypothesis to express the rate \eqref{eq:changeempRT-1} as a function of the one-particle distribution function only.
As for tumbling events, collisions change the empirical distribution; $f\left(\mathbf{r},\theta\right)$ is changed into
\begin{align}
    f\left(\mathbf{r},\theta\right)-\epsilon\delta\left(\mathbf{r}-\mathbf{r}_{1}\right)\delta\left(\theta-\theta_{1}\right)& - \epsilon\delta\left(\mathbf{r}-\mathbf{r}_{1}\right)\delta\left(\theta-\theta_{2}\right) \nonumber \\ & +\epsilon\delta\left(\mathbf{r}-\mathbf{r}_{1}\right)\delta\left(\theta-\theta'_{1}\right)+\epsilon\delta\left(\mathbf{r}-\mathbf{r}_{1}\right)\delta\left(\theta-\theta'_{2}\right).
\end{align}
The infinitesimal generator term accounting for collisions thus reads
\small
\begin{multline}
     G_{f,{\rm col}}[\phi]=\frac{1}{2\epsilon}\int d\theta_{1}d\theta_{2}d\theta'_{1}d\theta'_{2}d\mathbf{r}\ K(\theta_{2}-\theta_{1}) \\
     \times f(\mathbf{r},\theta_{1},t)f(\mathbf{r},\theta_{2},t) P_{\sigma}\left(\theta'_{1}-\Psi(\theta_{1},\theta_{2})\right)P_{\sigma}\left(\theta'_{2}-\Psi(\theta_{1},\theta_{2})\right) \left(\phi[\tilde{f}]-\phi[f] \right).
\end{multline}
\normalsize
where $\tilde{f}\left(\mathbf{r}_0,\theta,t\right)=f\left(\mathbf{r}_0,\theta,t\right)+\epsilon\delta(\mathbf{r}_{0}-\mathbf{r})\left(-\delta(\theta-\theta_{1})-\delta(\theta-\theta_{2})+\delta(\theta-\theta'_{1})+\delta(\theta-\theta'_{2})\right)$.
The large deviation Hamiltonian is deduced using \eqref{eq:HamiltoGen}
\begin{equation}
H_{BV}\left[f,p\right]=H_{\mathcal{T}}\left[f,p\right]+H_{{\rm col}}\left[f,p\right],\label{eq:Hamiltonian}
\end{equation}
where $H_{\mathcal{T}}$ is given by \eqref{eq:H_T} and the collision term
of the Hamiltonian reads
\begin{multline}
\begin{split}H_{{\rm col}}[f,p] & =\frac{1}{2}\int d\theta_{1}d\theta_{2}d\theta'_{1}d\theta'_{2}d\mathbf{r}\,K(\theta_{2}-\theta_{1})f(\mathbf{r},\theta_{1},t)f(\mathbf{r},\theta_{2},t)\times \end{split}
\\
P_{\sigma}\left(\theta'_{1}-\Psi(\theta_{1},\theta_{2})\right)P_{\sigma}\left(\theta'_{2}-\Psi(\theta_{1},\theta_{2})\right)\bigg\{ e^{-p(\mathbf{r},\theta_{1},t)-p(\mathbf{r},\theta_{2},t)+p(\mathbf{r},\theta'_{1},t)+p(\mathbf{r},\theta'_{2},t)}-1\bigg\}.\label{eq:H_C}
\end{multline}

Equation (\ref{eq:Hamiltonian}) along with  (\ref{eq:H_T},\ref{eq:H_C}) is the fluctuating kinetic theory for the Boltzmann-Vicsek model. The most probable evolution path satisfies the deterministic evolution
equation given by the Hamilton equation associated with $H_{BV}$
\begin{equation}
\partial_{t}f(\mathbf{r},\theta,t)=\frac{\delta H_{BV}}{\delta p(\mathbf{r},\theta,t)}[f,0]\,
=
-\mathbf{e}_{\theta}\cdot\nabla f(\mathbf{r},\theta,t)
+\mathcal{I}_{\rm col}[f](\mathbf{r},\theta,t)
\label{eq:hamilton-jacobi}
\end{equation}
which is the \emph{deterministic} Boltzmann--Vicsek equation \eqref{eq:rescaledboltzmannvicsek}.

Just as in the Run-and-Tumble case, collisions conserve locally the number of particles, and this is reflected in the fact that $\int d\mathbf{r}d\theta \,\delta H_{\rm col}/\delta p(\mathbf{r},\theta) =0$.
Furthermore, $H_{BV}$ is again non-quadratic in the conjugated momentum $p$. This means that dynamical large deviations
of the empirical distribution are non-Gaussian. Contrary to the tumbling
Hamiltonian \eqref{eq:H_t}, the Hamiltonian for collisions is quadratic
in $f$, because collisions considered in the Boltzmann--Vicsek dynamics
are binary. The Hamiltonian $H_{BV}$ share some similarities with the one derived in \cite{Bouchet2020} for the Boltzmann equation describing the dynamics of a passive dilute gas: quadraticity in the distribution function $f$ and exponential dependence on the conjugated momentum. At variance with that case, however, the collision rules of the Vicsek-Boltzmann model break time-reversal symmetry, and does not conserve momentum nor kinetic energy.

\section{Fluctuating hydrodynamics deeply in the ordered phase}
\label{sec:degondderiv}

In this section, we derive the fluctuating hydrodynamics from the Boltzmann--Vicsek LDP given by \eqref{eq:LDP_Boltzmann} and \eqref{eq:Hamiltonian}. This is done as a perturbative expansion in a small parameter, the Knudsen number 
$\alpha=\ell/L$, where $\ell$ is the mean free path and $L$ is the system size. $\alpha$ is also the time scale to reach a local equilibrium. As a first step, we introduce in section \ref{sec:spde} the macroscopic scaling with the Knudsen number, and associate a fluctuating Boltzmann--Vicsek equation with the Boltzmann--Vicsek LDP.  From there we adapt to the fluctuating case the framework developed in a deterministic setting in \cite{degond2008continuum}. In section \ref{subsec:Local-equilibria} we discuss the  local equilibria of the Boltzmann--Vicsek equation. These local equilibria are characterized by two slow modes: the density field, and the orientational order field. Then, in section \ref{subsec:Chapman=00003D002013Enskog-expansion-close} we obtain fluctuating hydrodynamic equations for these two slow modes. Further, we show that at leading order in the Knudsen number $\alpha$, the noise appearing in these hydrodynamic equations is Gaussian. 

\subsection{Macroscopic scaling and rephrasing of the Large Deviation Principle as a Stochastic PDE}
\label{sec:spde}

Since we are interested in large scales and long times, we introduce the macroscopic variables $\tilde{t}=\alpha t,\tilde{\mathbf{r}}=\alpha\mathbf{r}$, and define $\tilde{f}(\tilde{\mathbf{r}},\theta,\tilde{t})=f(\alpha^{-1}\tilde{\mathbf{r}},\theta,\alpha^{-1}\tilde{t})$, $\tilde{p}(\tilde{\mathbf{r}},\theta)=p(\alpha^{-1}\tilde{\mathbf{r}},\theta)$. Then
\begin{eqnarray}
&&H_{\mathcal{T}}[f,p]  = \frac{1}{\alpha} \tilde{H}_{\mathcal{T}}[\tilde{f},\tilde{p}]\\
&&H_{{\rm col}}[f,p]=\alpha^{-2}\tilde{H}_{\rm col}[\tilde{f},\tilde{p}],
\int d\mathbf{r} d\theta \,p\partial_{t}f =  \frac{1}{\alpha} \int d\tilde{\mathbf{r}}d\theta\,\tilde{p}\partial_{\tilde{t}}\tilde{f};
\end{eqnarray}
we remove the tildes in the following. 
Isolating the linear part in $p$ (which contributes to the deterministic evolution), the collision Hamiltonian can be written
\[
H_{{\rm col}}[f,p] =\int p(\mathbf{r},\theta) \mathcal{I}_{\rm col}[f](\mathbf{r},\theta) d\mathbf{r} d\theta + H_{{\rm col},{\rm stoch}},
\]
where $H_{{\rm col},{\rm stoch}}$ gathers all terms of order at least $2$ in $p$.
The empirical distribution then satisfies a large deviation principle with speed $\text{\ensuremath{\epsilon}}^{-1}$
and rate function 
\begin{equation}
J_{T}[f]=\frac{1}{\alpha^3}\int_{0}^{T}dt\,\underset{p}{{\rm sup}}\left\{ \int d\mathbf{r}d\theta\, p(\mathbf{r},\theta)\big(\alpha \partial_{t}f+\alpha e_\theta \cdot \nabla f-\mathcal{I}_{\rm col}[f]\big)-H_{{\rm col},{\rm stoch}}[f,p]\right\} .
\end{equation}
Notice the overall factor $\alpha^{-3}$ coming from the change of
time and space variables; the final time $T$ and the system size have also been rescaled.
Formally, this LDP can be recast as a stochastic PDE:
\begin{equation}
\alpha\big(\partial_t f +e_\theta \cdot \nabla f\big) -\mathcal{I}_{\rm col}[f] =\xi(\mathbf{r},\theta,t),
\label{eq:spde}
\end{equation}
where the left hand side is the deterministic Boltzmann-Vicsek equation, and the right hand side is a noise whose distribution satisfies the LDP 
\begin{equation}
\label{eq:LDPxi}
\mathbb{P}\left[\left\{ \xi(t)\right\} _{0\leq t<T}=\left\{ u(t)\right\} _{0\leq t<T}\right] \underset{\epsilon\downarrow 0}{\asymp}\exp\left(-\frac{1}{\epsilon\alpha^3}J_f[u]\right),
\end{equation}
with
\begin{equation}
\label{eq:Jfu}
J_f[u] = \int_0^T dt \int d\mathbf{r} \, \underset{p}{\rm sup}\left( \int_0^{2\pi} p u\, d\theta -H_{{\rm col},{\rm stoch}}[f,p]\right). 
\end{equation}
A consequence of \eqref{eq:LDPxi} is that we can express the variance of $\xi$ through the large deviation Hamiltonian
\begin{equation}
 \mathbb{E}\left[\xi\left[f\right]\left(\mathbf{r},\theta,t\right)\xi\left[f\right]\left(\mathbf{r}',\theta',t'\right)\right]=\epsilon\alpha^3 \frac{\delta^{2}H_{BV}}{\delta p(\mathbf{r},\theta,t)\delta p(\mathbf{r}',\theta',t')}\left[f,p=0\right].\label{eq:correlationxi}   
\end{equation}
Note that only $H_{{\rm col},{\rm stoch}}$ contributes to the second functional derivative of $H_{BV}$ with respect to $p$. 
From the original LDP, which is a statement on the probability distribution of $f$, to the above statement about the probability distribution of $\xi$, there is a change of variable, which should introduce a Jacobian factor. At the large deviations level however, this factor is negligible. 
We stress that the noise $\xi$ bears several features that are in stark contrast with the fluctuating kinetic theories derived in the weak-interaction limit~\cite{Dean_1996,bertini2015macroscopic}: it is multiplicative at the kinetic level (since its distribution depends on $f$), it is non-Gaussian (this is encoded in the fact that $H_{BV}$ is not quadratic in $p$), and it depends explicitly on the particle-particle interactions. 

Finally, the local conservation of the number of particles implies that whenever $\int u(\mathbf{r},\theta) d\theta \neq 0$,  $J_f[u]=+\infty$. Indeed, take any momentum field $p(\mathbf{r})$ independent of $\theta$; then $H_{{\rm col},{\rm stoch}}[f,p]=0$ and $\int u(\mathbf{r},\theta) p(\mathbf{r}) d\theta = p(\mathbf{r}) \int\! u d\theta \neq 0$. A good choice of $p(\mathbf{r})$ then makes the supremum in \eqref{eq:Jfu} as large as we wish. In the stochastic PDE \eqref{eq:spde}, this translates in the fact that the noise conserves the number of particles:
\begin{equation}
\int \xi(t,\mathbf{r},\theta) d\theta=0.
\label{eq:conservation_noise}
\end{equation}
Contrary to the case of passive dilute gases (where also momentum and energy are conserved), there is no other conservation law, reflecting the absence of these conservation laws at the level of the microscopic collisions.

\subsection{Local equilibria\label{subsec:Local-equilibria}}


We now discuss the local equilibria of the Boltzmann-Vicsek equation, i.e. distributions $f$ that make the collision kernel vanish $\mathcal{I}_{\rm col}[f]=0$. This is the crucial ingredient to derive the fluctuating hydrodynamics deeply in the ordered state because any initial condition should relax fast (over time scales of order $\alpha^{-1}$) towards these local equilibria.

For clarity, we choose the noise distribution $P_\sigma$ in the collision kernel \eqref{eq:I-coll} to be a Von Mises distribution $P_{\sigma}\left(\theta\right)=V_{s}(\theta)=\left(2\pi I_{0}\left(s\right)\right)^{-1}\exp\left(s\cos\theta\right)$, but any other choice for $P_\sigma$ with similar qualitative characteristics would be admissible.
This distribution has a circular variance $\sigma^{2}\left(s\right)=1-I_{1}(s)/I_{0}(s)$,
where $I_{j}$ is the modified Bessel function of order $j$. The variance
$\sigma^{2}$ is a decreasing function of $s$.

The local equilibria are the solutions of the integral equation  
\begin{equation}
\mathcal{I}_{\rm col}[f]\left(\theta\right)=0\iff f\left(\theta\right)=\frac{\iint d\theta_{1}d\theta_{2}\,f\left(\theta_{1}\right)f\left(\theta_{2}\right)K\left(\theta_{2}-\theta_{1}\right)V_{s}\left(\theta-\Psi\left(\theta_{1},\theta_{2}\right)\right)}{\int d\theta_{1}\,f\left(\theta_{1}\right)K\left(\theta_{1}-\theta\right)}.\label{eq:fixed-point-M}
\end{equation}
The homogeneous isotropic state ($f$ independent of the angle) is always a solution. This is the unique one when $\sigma>\sigma_c$: here the system is described by a single hydrodynamic variable, the density $\rho(\mathbf{r},t)$. We are interested in the regime $\sigma<\sigma_c$, when non isotropic local equilibria emerge. By rotation invariance, they are indexed by a local angle $\varphi(\mathbf{r},t)$; the local equilibria are then of the form $\rho(\mathbf{r},t)M_{\varphi\left(\mathbf{r},t\right)}$ and there are two hydrodynamic fields: $\rho$ and $\varphi$. By rotational symmetry, the dependence on $\varphi$ is simple: there exists a function $m$ such that $M_\varphi(\theta)=m(\theta-\varphi)$.

\begin{figure}[!t]
\centering \includegraphics[scale=0.25]{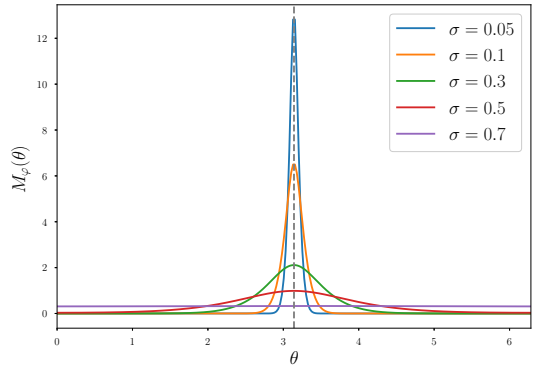} \caption{Profile of the local equilibria $M_{\varphi}(\theta)$ for different
values of $\sigma$ and $\varphi=\pi$. }
\label{fig:M_phi} 
\end{figure}
Although $M_{\varphi\left(\mathbf{r},t\right)}$ cannot be found analytically when $\sigma <\sigma_c$, finding it numerically is straightforward using the fixed-point formulation \eqref{eq:fixed-point-M}. We did this by implementing a fixed-point iteration method. For $\sigma>\sigma_c$, our algorithm correctly converges towards a constant solution, while for $\sigma<\sigma_c$, we obtain a solution for \eqref{eq:fixed-point-M} which carries a preferential orientation. Our numerical solutions for $M_{\varphi\left(\mathbf{r},t\right)}$ as a function of $\sigma$ is provided in Fig. \ref{fig:M_phi}. As it should be, the weaker this noise is, the narrower the local equilibrium $M_{\varphi}$ is around the local orientation $\varphi$. Obtaining $M_{\varphi\left(\mathbf{r},t\right)}$ with this method is very fast computationally, requiring only a few iterations unless $\sigma$ is set very close to $\sigma_c$. The value of $\sigma_c$ can be computed analytically \cite{Bertin2009}. To do so, one has to assess the linear stability of the collision operator $\mathcal{I}_{\rm col}$ linearized close to a uniform in angle distribution $f(\textbf{r},\theta,t)=\rho(\textbf{r},t)$. With the specific choice of a Von Mises distribution for the microscopical noise distribution $P_\sigma$, we have $\sigma_c = \sqrt{3}/3 \approx 0.58$.

\subsection{Chapman--Enskog expansion close to a local equilibrium\label{subsec:Chapman=00003D002013Enskog-expansion-close}}

In order to get the fluctuating hydrodynamics, we now want to compute evolution equations
for the density $\rho$ and the orientation field $\varphi$ that specify the local equilibria. To do so,  we
look for solutions to the kinetic equation \eqref{eq:spde}
as a Chapman--Enskog expansion close to a local equilibrium. This amounts to expand $f$ for small $\alpha$ as
\[
f\left(\mathbf{r},\theta,t\right)=\rho\left(\mathbf{r},t\right)M_{\varphi\left(\mathbf{r},t\right)}\left(\theta\right)+\alpha g\left(\mathbf{r},\theta,t\right)+\mathcal{O}\left(\alpha^{2}\right).
\]
At leading order in $\alpha$, we obtain from \eqref{eq:spde} that
\begin{equation}
 \left(\partial_{t}+\mathbf{e}_{\theta}\cdot\nabla\right)\left(\rho M_{\varphi}\right)- \rho L_{\varphi}[g] = \frac{1}{\alpha}\xi\left[\rho M_{\varphi}\right], \label{eq:chapman-kinetic}
\end{equation}
where $L_{\varphi}$ is the linearization of $\mathcal{I}_{\rm col}$ close
to $\rho M_{\varphi}$: 
\small
\[
L_{\varphi}[g](\theta)=\iint d\theta_{1}d\theta_{2}\,M_{\varphi}\left(\theta_{1}\right)g\left(\theta_{2}\right)K\left(\theta_{2}-\theta_{1}\right)\left\{ 2V_{s}\left(\theta-\Psi\left(\theta_{1},\theta_{2}\right)\right)-\delta\left(\theta-\theta_{1}\right)-\delta\left(\theta-\theta_{2}\right)\right\} .
\]
\normalsize
Classically, the Chapman--Enskog
expansion then proceeds integrating \eqref{eq:chapman-kinetic}
against conserved quantities, over the velocity variables (here the angle $\theta$). Each conserved quantity then yields
an evolution equation for a hydrodynamic mode. 

A difference with respect to the classical case arises here: we only have a single conserved quantity (density) and want to obtain evolution equations for both the density $\rho$ and the orientation field $\varphi$. Such problem was already discussed and solved in \cite{degond2008continuum} noting that, to obtain the evolution equation for the slow modes, we only need to integrate against quantities $\chi$ that are in the kernel of $L_{\varphi}^{\dagger}$, the adjoint operator of $L_{\varphi}$. The elements of the kernel of $L_{\varphi}^{\dagger}$ which do not correspond to conservation laws are known as Generalized Collisional Invariant (GCI).

\subsubsection{Equation for the density field.}
We observe that constants are in the kernel of $L_{\varphi}^{\dagger}$. Hence, integrating the fluctuating kinetic equation \eqref{eq:chapman-kinetic}
over $\theta$ yields the hydrodynamic equation for the density field
\begin{equation}
\partial_{t}\rho+c_{1}\nabla\cdot\left(\rho\mathbf{e}_{\varphi}\right)=0,\label{eq:density}
\end{equation}
where $c_{1}=\int d\theta\,\cos\left(\theta-\varphi\right)m\left(\theta-\varphi\right),$
and $\mathbf{e}_{\varphi}=\left(\cos\varphi,\sin\varphi\right)$ is
the orientational order field. We have used the density preserving property of the noise \eqref{eq:conservation_noise}.

\subsubsection{Equation for the orientational order field.}

In order to obtain a second hydrodynamic equation for the orientation field, we
need to find another element of $\ker L_{\varphi}^{\dagger}$ to integrate
 \eqref{eq:chapman-kinetic} against. In the classical kinetic
theory of passive gases, this second element is usually the velocity
variable \cite{bardos1991fluid}, which is a manifestation of momentum
conservation at the level of the kinetic equation. For active particles,
momentum conservation is broken and a GCI is needed.

Since $\mathcal{I}_{\rm col}\left[M_{\varphi}\right]=0$
for all $\varphi$, $\mathcal{I}_{\rm col}\left[M_{\varphi+\delta\varphi}\right]=0$ for any perturbation $\delta\varphi$.
This implies that not only $M_{\varphi}\in\ker L_{\varphi}$ but also
$\frac{\partial M_{\varphi}}{\partial\varphi}=-m'(\theta-\varphi) \in\ker L_{\varphi}$, which provides two elements in $\ker L_{\varphi}$ as soon as the system is locally ordered. Hence $\ker L^\dagger_{\varphi}$ is also two-dimensional, spanned by 
the constants and another element which we call $\psi_{\varphi}$: this is the GCI.

\begin{figure}[!t]
\centering \includegraphics[scale=0.25]{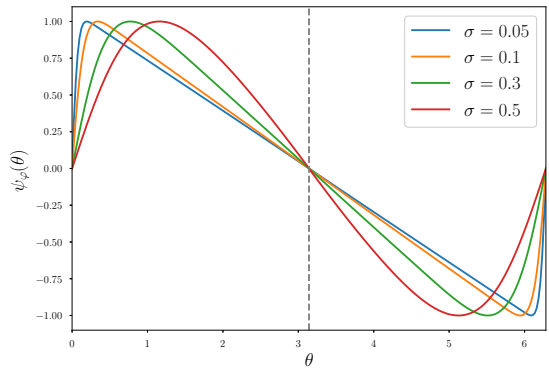} \caption{Profile of the GCI $\psi_{\varphi}(\theta)$ for different values
of $\sigma$ and $\varphi=\pi$. This quantity is defined whenever $\sigma < \sigma_c$.}
\label{fig:psi_phi} 
\end{figure}

 As it was the case for $M_\varphi$, $\psi_{\varphi}$ cannot be found analytically, but it can be determined numerically. 
In order to compute $\psi_{\varphi}$, we numerically solve the equation $L_{\varphi}^{\dagger}[\psi_{\varphi}]=0$ by discretizing
$\left[0,2\pi\right)$. Then, $L_{\varphi}^{\dagger}[\psi_{\varphi}]=0$
is a simple matrix equation that one can solve for $\psi_{\varphi}$. Observe that by rotational symmetry, the generalized collision invariant satisfies $\psi_{\varphi}\left(\theta\right)=-\psi_{-\varphi}\left(-\theta\right)$. In figure \ref{fig:psi_phi}, we plot $\psi_{\varphi}$ for $\varphi=\pi$
and for several values of $\sigma$.

Integrating \eqref{eq:chapman-kinetic} over $\theta$ and
against $\psi_{\varphi}$ and using that $L_{\varphi}^{\dagger}[\psi_{\varphi}]=0$
yields the hydrodynamic equation for the orientation field 
\begin{equation}
\label{eq:integralagainstpsi}
\alpha \int d\theta\,\psi_{\varphi}\left(\partial_{t}+e_\theta \cdot \nabla_{\mathbf{r}} \cdot\rho M_{\varphi}\right) = \int d\theta\,\psi_{\varphi}\xi\left[\rho M_{\varphi}\right],
\end{equation}
We see that for a smooth evolution of the orientation field $\varphi(\mathbf{r},t)$, the left hand side is of order $\alpha$, which corresponds to the noise $ \alpha \eta =  \int d\theta\,\psi_{\varphi}\xi\left[\rho M_{\varphi}\right]$ to be of order $\alpha$ as well. Contracting the probability distribution of $\xi$ given in (\ref{eq:spde}) and (\ref{eq:LDPxi}) to obtain the distribution of $\alpha\eta$, and expanding for small $\alpha$, it is easy to see that only the quadratic part of the distribution of $\alpha\eta$ contributes to leading order in $\alpha$. This is equivalent to saying that the noise becomes Gaussian in the hydrodynamic limit at leading order in $\alpha$. 


The explicit computation of the different terms in \eqref{eq:integralagainstpsi} yields the fluctuating hydrodynamic equation for the orientational order:
\begin{equation}
\rho\left(\partial_{t}\mathbf{e}_{\varphi}+c_{2}\mathbf{e}_{\varphi}\cdot\nabla\mathbf{e}_{\varphi}\right)+c_{3}\nabla_{\perp}\rho=\eta\mathbf{e}_{\perp} +\mathcal{O}(\alpha), \label{eq:orientation}
\end{equation}
where the $\mathcal{O}(\alpha)$ term represents the error committed in neglecting the higher order terms in the Chapman-Enskog expansion. In (\ref{eq:orientation}),  $\mathbf{e}_{\perp}=\mathbf{e}_{\varphi+\pi/2}$, $\nabla_{\perp}\rho=(\mathbf{e}_{\perp} \cdot\nabla \rho ) \mathbf{e}_{\perp}$
is the gradient of $\rho$ along the direction which is orthogonal to $\mathbf{e}_{\varphi}$ and 
\[
c_{4}=\frac{-1}{\int d\theta\,\psi_{\varphi}\left(\theta\right)m'\left(\theta-\varphi\right)},
\]
\[
c_{2}=-c_{4}\int d\theta\,\psi_{\varphi}\left(\theta\right)\cos\left(\theta-\varphi\right)m'\left(\theta-\varphi\right),
\]
and 
\[
c_{3}=c_{4}\int d\theta\,\psi_{\varphi}\left(\theta\right)\sin\left(\theta-\varphi\right)m\left(\theta-\varphi\right).
\]
Using the two-point correlations for $\xi$  \eqref{eq:correlationxi}, we can characterize the Gaussian noise $\eta$
\begin{equation}
\mathbb{E}\left[\eta\left(\mathbf{r},t\right)\eta\left(\mathbf{r}',t'\right)\right]=\alpha\epsilon C\rho{^{2}}\left(\mathbf{r},t\right)\delta\left(\mathbf{r}-\mathbf{r}'\right)\delta\left(t-t'\right)+\mathcal{O}(\alpha^2)\,,\label{eq:fluctuhydro_phi}
\end{equation}
and 
\[
C=c_{4}^{2}\left(C_{1}+C_{2}+C_{3}+C_{4}+C_{5}\right),
\]
with 
\small
\[
C_{1}=\int d\theta d\theta'\,\psi_{\varphi}\left(\theta\right)^{2}M_{\varphi}\left(\theta\right)M_{\varphi}\left(\theta'\right)K\left(\theta-\theta'\right),
\]
\[
C_{2}=\int d\theta d\theta'\,\psi_{\varphi}\left(\theta\right)\psi_{\varphi}\left(\theta'\right)M_{\varphi}\left(\theta\right)M_{\varphi}\left(\theta'\right)K\left(\theta-\theta'\right),
\]
\[
C_{3}=\int d\mathbf{\theta}d\theta'_{1}d\theta'_{2}\,\psi_{\varphi}\left(\theta\right)^{2}M_{\varphi}\left(\theta'_{1}\right)M_{\varphi}\left(\theta'_{2}\right)K\left(\theta'_{1}-\theta'_{2}\right)V_{s}\left(\theta-\Psi\left(\theta'_{1},\theta'_{2}\right)\right),
\]
\[
C_{4}=\int d\mathbf{\theta}d\theta' d\theta'_{1}d\theta'_{2}\,\psi_{\varphi}\left(\theta\right)\psi_{\varphi}\left(\theta'\right)M_{\varphi}\left(\theta'_{1}\right)M_{\varphi}\left(\theta'_{2}\right)K\left(\theta'_{1}-\theta'_{2}\right)V_{s}\left(\theta-\Psi\left(\theta'_{1},\theta'_{2}\right)\right)V_{s}\left(\theta'-\Psi\left(\theta'_{1},\theta'_{2}\right)\right),
\]
\[
C_{5}=-4\int d\theta d\theta'd\theta_{1}\,\psi_{\varphi}\left(\theta\right)\psi_{\varphi}\left(\theta'\right)M_{\varphi}\left(\theta\right)M_{\varphi}\left(\theta_1\right)K\left(\theta_{1}-\theta\right)V_{s}\left(\theta'-\Psi\left(\theta,\theta_{1}\right)\right).
\]
\normalsize
Although it is not apparent from the above expressions, we have checked numerically that $C$ is positive and an increasing function
of $\sigma$, as expected.

The structure of the fluctuating equation for the local orientation
field \eqref{eq:orientation} is not usual. In relation with the
lack of momentum conservation, \eqref{eq:orientation} contains
a noise term but no diffusive terms. These would give corrections at $\mathcal{O}(\alpha)$ in (\ref{eq:orientation}) and we expect that they can be obtained by similar lines as in \cite{degond2010diffusion} where they were derived for the Vicsek model within the weak-interaction limit; we leave this for future investigations. We should however observe that (\ref{eq:orientation}) allows already to obtain the path probability for $\mathbf{e}_\varphi$ to first order in $\alpha$, which is the central result of this Section.

The two main novelties of our results are the following. First, we obtain the hydrodynamics of self-propelled aligning particles for dilute systems, deeply in the ordered phase, which was not even known at the deterministic level, since the results of \cite{degond2008continuum} were derived in the weak-interaction limit. Second, we obtained also the hydrodynamics at the fluctuating level. The fact that we work in the dilute regime implies that the hydrodynamic noise variance is proportional to $\rho^2$, and that the noise depends explicitly on the collision rules (interactions) among particles. Both of these facts are at variance with the fluctuating hydrodynamics obtained in the weak-interaction regime -- the regime where one particle interacts with many others and noise comes from angular diffusion rather than collisions~\cite{bertin2013mesoscopic}. For the sake of clarity, we only consider noise that stems from collisions. Adding angular diffusion to the model would slightly modify the kinetic equation by adding a diffusion term as well as a Dean-like noise. At the deterministic hydrodynamics level, this would result in a modification of the linearized collision kernel $L_\varphi$, the shape of the local equilibrium $M_\varphi$, and thus of the hydrodynamic coefficients. At the fluctuating level, there would be a new noise term whose variance is proportional to $\rho$ instead of $\rho^2$. We expect this noise to be the dominant one at low densities.

\section{Fluctuating hydrodynamics close to order-disorder transition}
\label{sec:hydrophasetransition}
In this last section we derive the fluctuating hydrodynamics with the same starting point -the fluctuating Boltzmann-Vicsek equation- but in a different regime: close to the order-disorder transition. We follow the route first introduced in \cite{Bertin2006}. This relies on an expansion close to the instability threshold $\sigma_c$ obtained via a moment expansion and a closure. This method has since been widely used in the literature~\cite{chate2020dry}. Yet it should be noticed that its quantitative regime of validity is unclear: the order-disorder transition is generically first order in the Vicsek model and $\sigma_c$ is well defined only at mean-field level~\cite{gregoire2004onset,martin2021fluctuation}. 

The fluctuating hydrodynamics derived close to the instability threshold  was previously obtained (for nematic systems) adding a Dean-like noise to the Boltzmann-Vicsek kinetic equation \cite{bertin2013mesoscopic}. Here, starting from our fluctuating kinetic theory (\ref{eq:LDP_Boltzmann}, \ref{eq:Hamiltonian}), we derive the noise term close to the instability threshold and compare to the results of section \ref{sec:degondderiv}, which are valid deeply in the ordered regime. 


The derivation of the deterministic hydrodynamics can be found in \cite{Bertin2006, Bertin2014} and we only sketch it here in section \ref{sec:moment_expansion}. In section \ref{sec:BGL}, we use the same Ansatz as in \cite{Bertin2006} to show that a Gaussian approximation for the noise is justified, and to find the fluctuating evolution equations for the slow hydrodynamic modes. We then check that there exists a scaling regime (in terms of microscopic parameters) in which both the hydrodynamic limit and the noise terms are controlled. 
Last, in section \ref{sec:connection}, we show that both hydrodynamic derivations from sections \ref{sec:degondderiv} and \ref{sec:hydrophasetransition} can be connected at leading order.

\subsection{Moment expansion}
\label{sec:moment_expansion}
Our starting point is the same as in section \ref{sec:degondderiv}: after rescaling time and space with the Knudsen number $\alpha$ we work with the fluctuating Boltzmann-Vicsek equation \eqref{eq:spde} together with the LDP for the noise \eqref{eq:LDPxi}.

As customary~\cite{Bertin2014}, we introduce the complex derivatives  $\nabla = \partial_x + i \ \partial_y, \nabla^{\star} = \partial_x - i \ \partial_y, \Delta = \nabla \nabla^{\star}$ and the following notations for the Fourier transforms
\begin{align}
f_k(\mathbf{r},t) = \int_{0}^{2\pi} d\theta \, e^{i k \theta}f(\mathbf{r},\theta,t),
\end{align}
and a similar notation for $\xi$. 
Taking the Fourier transform of \eqref{eq:spde} we thus obtain
\begin{equation}
\label{eq:moment_expansion}
\alpha\big( \partial_t f_k + \frac{1}{2} (\nabla f_{k-1}+\nabla^{\star} f_{k+1}) \big) = \sum_{q=-\infty}^{+\infty} (P_k I_{k,q}-I_{0,q})f_q f_{k-q}+ \xi_k
\end{equation}
with 
\begin{align}
    P_k(\sigma) = \int_{0}^{2 \pi} d\theta \, P_{\sigma}(\theta) e^{i k \theta},\qquad I_{k,q} = \frac{1}{2 \pi} \int_{0}^{2 \pi} d \Delta \,  K(\Delta) e^{-i q \Delta + i k H(\Delta)} \,.
\end{align}
The noise terms $\xi_k$ are Fourier transforms of $\xi$, which is non-Gaussian, and in particular satisfies a non-quadratic LDP given by \eqref{eq:LDPxi} and \eqref{eq:Jfu}. However, similarly to the case of section \ref{sec:degondderiv} we shall argue below that we are actually interested in the small $\xi$ limit, in which the large deviation function can be considered quadratic, and $\xi_k$ be approximated by a Gaussian noise. 
  
\subsection{Boltzmann--Ginzburg--Landau scaling}
\label{sec:BGL}
We now aim at finding the evolution equation for the Fourier modes of the distribution function in the hydrodynamic limit, where usually only the first few Fourier modes matter, by truncating the infinite hierarchy of stochastic partial differential equations in \eqref{eq:moment_expansion}. Since mass is a conserved quantity, we already know that the density field $\rho(\mathbf{r},t)=f_0(\mathbf{r},t)$ is a relevant hydrodynamic field. Assuming a scaling Ansatz similar to the one used in \cite{Bertin2006} for polar particles, we will show that the polarity field $f_1(\mathbf r,t)$ is the second slow field of the problem while the other modes $f_{k>1}$ are fast fields.

The equations for the first $k$ modes, $k \in \{0,1,2\}$, read
\begin{align}
    & \partial_t f_0 + \frac{1}{2} \text{Re}{\nabla^{\star} f_1} = 0, \label{eq:f0}\\
    & \alpha \big(\partial_t f_1 + \frac{1}{2}(\nabla f_0 + \nabla^{\star} f_2) \big) =  \mu_1[\rho] f_1 + (J_{1,2}+J_{1,-1}) f_1^{\star} f_2 + ... + \xi_1, \label{eq:f1}\\
       &  \alpha \big(\partial_t f_2 + \frac{1}{2} (\nabla f_1 + \nabla^{\star}f_3) \big)= \mu_2[\rho] f_2 + J_{2,1} f_1^2 +... + \xi_2, 
\label{eq:hierachy}
\end{align}
where ... denotes the other terms coming from the collision kernel in  \eqref{eq:moment_expansion}, $J_{k,q} = P_k(\sigma) I_{k,q}-I_{0,q}$ and $\mu_{k}[\rho] = \left( J_{k,0} + J_{k,k} \right) \rho_0$.
If noises in the hierarchy of equations  \eqref{eq:f0}--\eqref{eq:hierachy} are switched off, the system of equations admits a solution $\{f_0(\textbf{r},t) = \rho_0, f_{k>0} = 0 \}$, which corresponds to the homogeneous disordered state. It turns out that, at fixed density $\rho_0$, this solution is linearly stable only when $\sigma$ is greater than a threshold value $\sigma_c$, in which case all the $\mu_k[\rho_0]$'s are negative. Below $\sigma_c$, an instability of the disordered state is triggered because $\mu_1[\rho_0]$ changes sign while the $\mu_{k>1}[\rho_0]$ do not ~\cite{Bertin2014}. This is the regime in which we work in the rest of this section.

In order to decouple the evolution of the slow hydrodynamic modes from the fast ones, we work at the onset of instability. Indeed by tuning $\sigma$, it is possible to control at will the size of $\mu_1[\rho]$ (as it changes sign continuously) and we choose to work in a scaling regime such that $\mu_1[\rho] = \alpha^2 \mu_1'[\rho]$. We moreover assume an Ansatz à la Boltzmann--Ginzburg--Landau in this scaling regime which reads 
\begin{equation}
\label{eq:GLscaling_polar}
        f_0 = \rho_0 + \alpha \delta \rho , f_{1} = \alpha f_1', f_{k > 1} = \alpha^2 f_k'.
\end{equation}
To alleviate notations, we drop the primes. Introducing this scaling into  (\ref{eq:f1}--\ref{eq:hierachy}), we see that we are actually interested in a regime where the noise terms $\xi_k$, and hence $\xi$ itself, are small. By a similar reasoning as in section \ref{sec:spde}, we conclude that the large deviation rate function in \eqref{eq:LDPxi} has to be considered only for small values of the variable $u$ and it is legitimate to use a Gaussian approximation. As a consequence, we shall from now on use a Gaussian approximation for the noise $\xi$.  In particular, it is characterized by its first two moments, which can be computed, for any integers $k,l$
\begin{align}
\label{eq:esperance_mode_k}
         \mathbb{E}[ \xi_k(\mathbf{r},t) ] = &\int_{0}^{2\pi} d\theta \, e^{i k \theta} \mathbb{E}[ \xi(\mathbf{r},\theta,t) ] = 0, \\
\label{eq:variance_mode_k}        
         \mathbb{E}[ \xi_k(\mathbf{r},t)\xi_l^{\star}(\mathbf{r}',t') ] =& \int_{0}^{2\pi} d\theta  d\theta' \, e^{i k \theta}e^{-i l \theta'} \mathbb{E}[ \xi(\mathbf{r},\theta,t)\xi(\mathbf{r}',\theta',t') ]\\ = &\ \ \ \epsilon \alpha^3  \mathcal{V}_{\rm col}  \delta(\mathbf{r}-\mathbf{r}')\delta(t-t'),
\end{align}
with
\begin{align}
\label{eq:variance}
        & \mathcal{V}_{\rm col} = \sum_{q = -\infty}^{\infty} \nu_{k,l,q}(\sigma) \ f_{k-l+q}(\mathbf{r},t)f^{\star}_{q}(\mathbf{r},t), \\
        & \nu_{k,l,q}(\sigma) = \frac{1}{2} \bigg[I_{0,k-l+q}+I_{0,k+q}+I_{0,q-l}+I_{0,q} -2 \bigg(P_k(\sigma ) I_{k,k+q}+P_{-l}(\sigma ) I_{-l,q-l}\bigg) \\
        &  -2 \bigg(P_k(\sigma ) I_{k,k-l+q}+P_{-l}(\sigma ) I_{-l,k-l+q}\bigg)+2  \bigg(P_{k-l}(\sigma )+P_k(\sigma ) P_{-l}(\sigma )\bigg) I_{k-l,k-l+q} \bigg] \,.
\end{align}
If we look at self-correlations, we have
\begin{equation}
        \mathbb{E}[ \xi_k(\mathbf{r},t)\xi_k^{\star}(\mathbf{r}',t') ]  = \delta(\mathbf{r}-\mathbf{r}')\delta(t-t') \sum_{q = 0}^{\infty} \gamma_{k,q}(\sigma) \ \abs{f_{q}(\mathbf{r},t)}^2,
\end{equation}
with
\begin{align}
\label{eq:def_alpha}
        \gamma_{k,q}(\sigma) & = \nu_{k,k,-q}(\sigma)+\nu_{k,k,q}(\sigma) \ \text{if $q \neq 0$}, \\
        \gamma_{k,0}(\sigma) & = \nu_{k,k,0}(\sigma) \,.
\end{align}

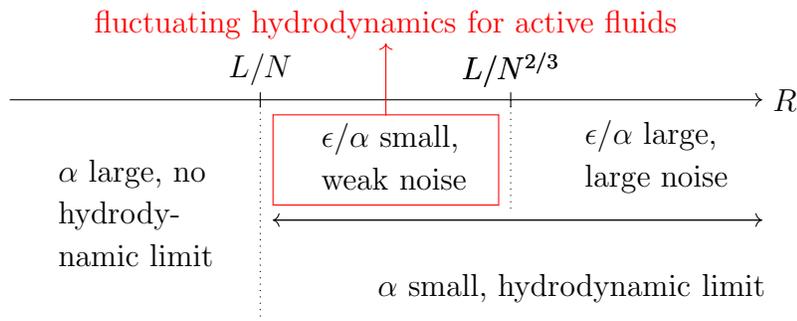
\begin{figure}[!t]
\centering
\begin{tikzpicture}[baseline=-.2cm,decoration={markings,mark=at position 1.0 with {\arrow{>}}}]
    \coordinate (N1)     at  (0,0);
    \coordinate (N2)     at  (10,0);
    \coordinate (N3)     at  (3.33,0);
    \coordinate (N31)     at  (3.33,0.1);
    \coordinate (N32)     at  (3.33,-0.1);
    \coordinate (N33)     at  (3.33,0.4);
    \coordinate (N4)     at  (3.33,-3);
    \coordinate (N5)     at  (6.66,0);
    \coordinate (N51)     at  (6.66,0.1);
    \coordinate (N52)     at  (6.66,-0.1);
    \coordinate (N53)     at  (6.66,0.4);
    \coordinate (N6)     at  (6.66,-1.5);
    \coordinate (N71)     at  (0.5,-1.5);
    \coordinate (N72)     at  (4.75,-2.5);
    \coordinate (N81)     at  (3.5,-1.6);
    \coordinate (N82)     at  (10.0,-1.6);
    \coordinate (N91)     at  (4.0,-0.75);
    \coordinate (N92)     at  (7.5,-0.75);
    \coordinate (N93)     at  (5.0,-0.21);
    \coordinate (N94)     at  (5.0,0.75);
    \coordinate (N95)     at  (5.0,1.0);
    \draw[postaction={decorate}]  (N1)   -- (N2);
    \draw[postaction={decorate}]  (N81)   -- (N82);
    \draw[postaction={decorate}]  (N82)   -- (N81);
    \draw[postaction={decorate},red]  (N93)   -- (N94);
    \draw [dotted] (N3) -- (N4);
    \draw (N31) -- (N32);
    \draw [dotted] (N5) -- (N6);
    \draw (N51) -- (N52);
    \path (N2) node [right] {$R$};
    \path (N33) node {$L/N$};
    \path (N53) node {$L/N^{2/3}$};
    \path (N53) node {$L/N^{2/3}$};
    \path (N71) node [right,text width=2.5cm] {$\alpha$ large, no hydrodynamic limit};
    \path (N72) node [right] {$\alpha$ small, hydrodynamic limit};
    \path (N91) node [right,text width=2cm] {$\epsilon/\alpha$ small, weak noise};
    \path (N92) node [right,text width=2cm] {$\epsilon/\alpha$ large, large noise};
    \path (N95) node [red] {fluctuating hydrodynamics for active fluids};
    \draw[draw=red] (3.5,-0.2) rectangle ++(3.0,-1.2);
\end{tikzpicture}
\caption{Scaling limit}
\label{fig:scaling_region}
\end{figure}

Now, given the fact that $\mu_2[\rho] < 0$ close to the (first) instability line, we assume $\partial_t f_2 \approx 0$. We then collect only the terms of lowest order in $\alpha$ in the equations for the modes 0, 1, 2, including noise terms and their cross-correlations. This enslaving procedure notably generates, in the equation for the mode $f_1$, the stochastic terms $\alpha \nabla^{\star} \xi_2$ and $f_1 \xi_2$. These terms are however of order $\mathcal{O}(\alpha^{5/2} \varepsilon^{1/2})$ and are thus less relevant than the noise term $\xi_1$, which is of order $\mathcal{O}(\alpha^{3/2} \varepsilon^{1/2})$. The same happens for the cross-correlations terms, which turn out to be of higher order than $\mathcal{O}(\alpha^{3/2} \varepsilon^{1/2})$. We thus end up with the following equations 
\begin{align}
\label{eq:noisy_hydro_polar}
    & \partial_t \delta \rho + \Re{ \nabla f_1^{\star} } = 0, \\
    & \partial_t f_1 = - \frac{1}{2} \nabla \delta \rho + \alpha \bigg\{ \mu_1 f_1 - \beta f_1 \abs{f_1}^2 + \nu \Delta f_1 + \kappa_1 f_1 \nabla^{\star} f_1 + \kappa_2 f_1^{\star} \nabla f_1 \bigg\} + \eta_1,
\end{align}
whose deterministic part is derived in \cite{Bertin2014}. $\eta_1$ is a Gaussian white noise whose correlations read at lowest order in $\alpha$
\begin{align}
\label{eq:effective_hydro_noise_polar}
    & \mathbb{E}[ \eta_1(\textbf{r},t) \eta_1^{\star}(\textbf{r}',t') ]= \frac{\epsilon}{\alpha} \gamma_{1,0} \rho_0^2 \ \delta(t-t') \delta(\textbf{r}-\textbf{r}'), \\
    & \mathbb{E}[ \eta_1(\textbf{r},t) \eta_1(\textbf{r}',t') ] = 0 .
\end{align}
$\gamma_{1,0}$ is defined in \eqref{eq:def_alpha} and can be checked to be positive once a specific form for the collision kernel $K$ has been chosen. The other coefficients are given by
\begin{align}
    \nu = \frac{1}{4 \abs{\mu_2}}, \kappa_1 = & \frac{1}{\mu_2}(P_2 I_{2,1}-I_{0,1}),
    \kappa_2 = \frac{1}{2 \mu_2} \bigg[P_1 (I_{1,-1}+I_{1,2}) - I_{0,-1} - I_{0,2} \bigg], \\
    & \beta = \frac{P_2 I_{2,1}-I_{0,1}}{\mu_2} \bigg[P_1 (I_{1,-1}+I_{1,2}) - I_{0,-1} - I_{0,2} \bigg] \,.
\end{align}
We work in the scaling limit where $\alpha \to 0$. However, in this scaling limit, we additionally want the strength of hydrodynamic noise to be small, i.e. $\epsilon/\alpha \to 0$. It turns out that
\begin{align}
    & \alpha = \frac{L}{R N} \to 0 \implies \frac{L}{N} \ll R, \\
    & \frac{\epsilon}{\alpha} = \frac{N^2 R^3}{L^3} \to 0 \implies R \ll \frac{L}{N^{2/3}},
\end{align}
where $L,R,N$ were defined in section \ref{subsec:Pmodel}. As $N$ is large, $N \gg N^{2/3}$ and there exists a scaling region in which both $\alpha$ and $\epsilon/\alpha$ can be arbitrary small, as shown in figure \ref{fig:scaling_region}. This is the scaling limit we choose.

The Langevin equations (\ref{eq:noisy_hydro_polar}-\ref{eq:effective_hydro_noise_polar}) are consistent at the deterministic level with the one studied in \cite{Toner1998, Bertin2006, Bertin2014}. However, as it is the case in section \ref{sec:degondderiv}, the noise acting on the orientation field differs from the ones previously considered in the literature for two reasons. First, it is proportional to $\rho_0^2$; second, it explicitly depends on the particle interactions via $\gamma_{1,0}$. Both these facts are generically expected in the fluctuating hydrodynamic description of dilute active systems.\\

It should be further observed that different sources of noises add up linearly within this framework. In particular, if we had included translational diffusion of active particles, this would generate at the hydrodynamic level the deterministic diffusion terms which are discussed in \cite{Bertin2014} and noise terms $\eta_D$ respecting mass conservation in the equation for $\delta \rho$ and $f_1$, with correlations $\mathbb{E}[\eta_D(\textbf{r},t)\eta_D(\textbf{r}',t')] \propto \delta(t-t') \nabla^2\delta(\textbf{r}-\textbf{r}')$. 
Similarly to the deeply in the ordered phase case (section 3), adding angular diffusion or run-and-tumble dynamics to the particle dynamics would simply  modify the hydrodynamics coefficients at the deterministic level. At the fluctuating level, it would create a new noise term in the equation for $f_1$ whose variance is proportional to $\rho$.  

\subsection{Connection between hydrodynamics equation deeply in the ordered phase and close to the phase transition}
\label{sec:connection}

We conclude comparing the hydrodynamics obtained deeply in the ordered phase (section \ref{sec:degondderiv}) with the one close to the instability threshold (section \ref{sec:hydrophasetransition}). Following \cite{mahault2018}, we express \eqref{eq:noisy_hydro_polar} in terms of density and polarity fields, by defining $f_1(\mathbf{r},t) = p_x+i p_y, \text{where} \ \mathbf{p} = (p_x,p_y)$ is the polarity field. They read 
\begin{align}
& \partial_t \delta \rho + \nabla \cdot \mathbf{p} = 0, \\
& \partial_{t}\mathbf{p}+\lambda_{1}\left(\mathbf{p}\cdot\nabla\right)\mathbf{p}+\lambda_{2}\left(\nabla\cdot\mathbf{p}\right) \mathbf{p}-\frac{\lambda_{2}}{2}\nabla\left(\mathbf{p}^{2}\right)=\left(a-b\mathbf{p}^{2}\right)\mathbf{p} - c_3 \nabla \delta \rho+D_{T}\Delta\mathbf{p} + \mathbf{\eta},\label{eq:ogTT}
\end{align}
where
\begin{align}
\label{eq:hydroTonerTu}
    & \lambda_1 = \alpha(\kappa_1+\kappa_2), \lambda_2 = \alpha(\kappa_1-\kappa_2), a = \alpha \mu_1, b = \alpha \xi, c_3 =\frac{1}{2}, D_T = \alpha \nu,
\end{align}
and $\mathbf{\eta} = \left( \eta _i \right)_{i=1,2} $ is an isotropic Gaussian white noise whose correlations read
\begin{equation}
    \mathbb{E}[ \eta_i(\mathbf{r},t) \eta_j(\mathbf{r}',t') ] = \frac{1}{2} \frac{\epsilon}{\alpha} \delta_{ij}  \gamma_{1,0} \rho_0^2  \delta(t-t') \delta(\mathbf{r}-\mathbf{r}').
\end{equation}
If one further assumes that \eqref{eq:ogTT} holds far away for the transition to collective motion, one can wonder how the evolution equation for the polarity field is affected. Assuming the norm of the polarity field to be fixed to $p_0=\sqrt{a/b}$, which is reasonable deep in the ordered phase, we look for solutions under the form $\mathbf{p}=p_0 \mathbf{e}_{\varphi}$, where $\mathbf{e}_{\varphi}$ is a unit 2d-vector parametrized by the angle $\varphi$. Projecting \eqref{eq:hydroTonerTu} onto $\mathbf{e}_{\perp}=\mathbf{e}_{\varphi+\pi/2}$ yields
\begin{equation}
    p_0 \partial_t \mathbf{e}_{\varphi} + \lambda_1 p_0^2  \bigg[ ( \mathbf{e}_{\varphi} \cdot \nabla) \mathbf{e}_{\varphi} \bigg] = - c_3 \nabla_{\perp} \delta \rho + D_T p_0 ( \mathbf{e}_{\perp} \cdot \Delta \mathbf{e}_\varphi )\mathbf{e}_{\perp} + ( \mathbf{e}_{\perp} \cdot \mathbf{\eta})\mathbf{e}_{\perp}
    \label{eq:polarordertonertu}
\end{equation}
We recognize in  \eqref{eq:polarordertonertu} all the terms present in \eqref{eq:orientation}.
However these two equations differ for two reasons. First, the dependence of the parameters entering the hydrodynamic description on the microscopic ones differs in the two cases, both at deterministic and fluctuating level. Second, the Laplacian term in \eqref{eq:polarordertonertu} is of the same order as transport terms, while these Laplacian terms were subdominant (and hence neglected) in \eqref{eq:orientation}. 

\section{Conclusions}
We focused on active matter systems where polar alignment is the dominant interaction in the dilute regime. Within this framework, we have extended the Boltzmann deterministic kinetic theory to its fluctuating counterpart. This is best described through a large deviation theory formalism, given that fluctuations in the kinetic theory are not Gaussian. The large deviation Hamiltonian associated with it is given in \eqref{eq:Hamiltonian}, \eqref{eq:H_C}. Our fluctuating Boltzmann-Vicsek equation has the same regime of validity as the original Boltzmann-Vicsek equation: $\epsilon\ll1$, where $\epsilon^{-1}=\rho_0 \ell^2$ is the number of particles in an area equal to the square of the mean free path $\ell$. 

We have then derived the associated fluctuating hydrodynamics in two different regimes of parameters. First, deeply in the ordered phase, our final result is \eqref{eq:density}, \eqref{eq:orientation}, \eqref{eq:fluctuhydro_phi}, which allows to obtain the path probability of the density and the orientational field to leading order in the Knudsen number $\alpha=\ell/L$, where $L$ is a macroscopic length-scale (e.g. the size of the system). In this regime and for dilute systems, even the derivation of the deterministic hydrodynamics was not known.
We stopped the perturbative expansion at leading order in $\alpha$, which corresponds to neglecting diffusive terms, but the same technique could be employed to obtained them, along the lines of the computations previously done in the weak-interactions regime~\cite{degond2008continuum}. Second, we have derived the fluctuating hydrodynamics close to the transition between order and disorder using a moment expansion and a closure of the hierarchy, as widely employed in the active matter community~\cite{chate2019dry}.

The derivation of the hydrodynamic noise in the dilute regime differs in two important aspects from the one obtained in the weak-interactions regime~\cite{Dean_1996,bertin2013mesoscopic}. First, it depends explicitly on the particle interactions and, reflecting the binary nature of the collisions, its variance is quadratic in the density. 

We conclude with three remarks. First, we have presented results on polar particles with polar aligning interactions, but we expect that these can be generalized to polar particles with nematic interactions or to fully active nematic systems. Second, in most real systems stochasticity at hydrodynamic level can originate both from interactions and from single-particle diffusion; our theory is linear -- and if single particle diffusion is present, it just adds up at hydrodynamic level. Lastly, while our derivation of the hydrodynamic theory deeply in the ordered state can be considered controlled from a mathematical viewpoint, it should be noted that our results assume a small noise on top of a smooth evolution at the hydrodynamic level: our scaling hypothesis might break down in the presence of shocks. The analysis of large deviations in their presence is a much harder problem, whose understanding is so far limited only to few examples~\cite{bodineau2005distribution}, and progress along these lines would certainly require the analysis of dissipative terms at hydrodynamic level.

\ack 
OF, MB,  CN, JB warmly thank Freddy Bouchet for insightful discussions concerning his work~\cite{Bouchet2020} that inspired the present one. JB acknowledges many discussions with C\'edric Bernardin and Rapha\"el Ch\'etrite. This work has been supported by the project RETENU ANR-20-CE40-0005-01 of the French National Research Agency (ANR). OF, MB,  CN, JB thank the anonymous referee for their useful recommendations, which helped us to improve the first version of the manuscript.

\bibliographystyle{iopart-num.bst}
\bibliography{mainarticle_C+O+J}
\end{document}